\documentclass{ws-ijmpa}
\usepackage[super,compress]{cite}
\usepackage{graphicx}
\usepackage{graphicx}
\usepackage{dcolumn}
\usepackage{bm}
\usepackage{epstopdf}
\newcommand{\hs}{\hspace*{0.5cm}}

\newcommand{\be}{\begin{equation}}
\newcommand{\ee}{\end{equation}}
\newcommand{\bea}{\begin{eqnarray}}
\newcommand{\eea}{\end{eqnarray}}
\newcommand{\ben}{\begin{enumerate}}
\newcommand{\een}{\end{enumerate}}
\newcommand{\bde}{\begin{widetext}}
\newcommand{\ede}{\end{widetext}}
\newcommand{\nn}{\nonumber}
\newcommand{\crn}{\nonumber \\}

\newcommand{\la}{\lambda}

\newcommand{\om}{\omega}

\newcommand{\fr}{\frac}
\newcommand{\bc}{\begin{center}}
\newcommand{\ec}{\end{center}}

\newcommand{\La}{\Lambda}
\newcommand{\si}{\sigma}

\begin{document}
%
\catchline{}{}{}{}{}

\title{NEUTRINO MIXING WITH NON-ZERO $\theta_{13}$ AND CP VIOLATION\\ 
IN THE 3-3-1 MODEL BASED ON $S_4$ FLAVOR SYMMETRY}

\author{VO VAN VIEN }

\address{Department of Physics, Tay Nguyen University, 567 Le
Duan, Buon Ma Thuot, Vietnam\\
wvienk16@gmail.com}

\author{HOANG  NGOC  LONG}

\address{Institute of Physics,
VAST, 10 Dao Tan, Ba Dinh, Hanoi, Vietnam\\
hnlong@iop.vast.ac.vn}

\author{DINH PHAN KHOI }

\address{Department of Physics, Vinh University, 182 Le Duan, Vinh City, Nghe An, Vietnam\\
khoidp@vinhuni.edu.vn}

\maketitle
\begin{history}
\received{Day Month Year}
\revised{Day Month Year}
\end{history}

\begin{abstract}
The 3-3-1 model proposed in 2011 based on discrete symmetry $S_4$
responsible for the neutrino and quark masses is updated, in which
the non-zero $\theta_{13}$ is focused. Neutrino masses and mixings
are consistent with the most recent data on neutrino oscillations
without perturbation. The new feature is adding a new  $SU(3)_L$
anti-sextet lying in
 doublet under $S_4$ which can result the non-zero $\theta_{13}$ without perturbation,
and consequently, the number of Higgs multiplets required is less than
those of other models based on non-Abelian discrete symmetries and the 3-3-1 models. The exact tribimaximal form obtained with the breaking $S_4 \rightarrow Z_3$ in charged lepton sector and $S_4 \rightarrow \mathcal{K}$
in neutrino sector. If both breakings
$S_4\rightarrow \mathcal{K}$ and $\mathcal{K} \rightarrow Z_2$ are taken place in neutrino sector,
the realistic neutrino spectrum is obtained without perturbation. The upper bound on neutrino
mass and the effective mass governing neutrinoless double beta decay at the tree level are presented.
The model predicts the Dirac CP violation phase  $\delta=292.45^\circ$ in the normal spectrum (with
$\theta_{23}\neq \frac{\pi}{4}$) and $\delta=303.14^\circ$ in the inverted spectrum.

\end{abstract}

\ccode{PACS numbers:14.60.Pq, 14.60.St, 11.30.Hv, 11.30. Er, 12.60.-i}


\section{\label{intro}Introduction}

Nowadays, particle physicists are attracted by two exciting
subjects: Higgs and neutrino physics. The neutrino mass and mixing
are the first evidence of beyond Standard Model physics. Many
experiments show that neutrinos have tiny masses and their mixing
is sill mysterious\cite{altar1, altar2} .
The tri-bimaximal form for explaining the lepton mixing scheme was first
 proposed by Harrison-Perkins-Scott (HPS), which
apart from the phase redefinitions, is given by \cite{hps1,hps2,hps3,hps4}
\begin{eqnarray}
U_{\mathrm{HPS}}=\left(
\begin{array}{ccc}
\frac{2}{\sqrt{6}}       &\frac{1}{\sqrt{3}}  &0\\
-\frac{1}{\sqrt{6}}      &\frac{1}{\sqrt{3}}  &\frac{1}{\sqrt{2}}\\
-\frac{1}{\sqrt{6}}      &\frac{1}{\sqrt{3}}  &-\frac{1}{\sqrt{2}}
\end{array}\right),\label{Uhps}
\end{eqnarray}
can be considered  as a good approximation for the recent
neutrino experimental data.

The most recent data are a clear sign of
rather large value $\theta_{13}$ \cite{smirnov} . The data in the Particle Data Group PDG2014 \cite{PDG2014} imply:
\bea
&&\sin^2(2\theta_{12})=0.846\pm 0.021 , \,\,\,\, \sin^2(2\theta_{23})=0.999^{+0.001}_{-0.018}, \crn
&&\sin^2(2\theta_{13})=(9.3\pm0.8)\times 10^{-2},\,\,\, \Delta m^2_{21}=(7.53\pm 0.18)\times 10^{-5} \mathrm{eV}^2,\crn
&&   \Delta m^2_{32}=(2.44\pm0.06)\times
10^{-3}\mathrm{eV}^2 ,\,\, (\mathrm{Normal \,\,hierarchy}),\label{PDG14N}\\
&&\sin^2(2\theta_{12})=0.846\pm 0.021 , \,\,\,\, \sin^2(2\theta_{23})=1.000^{+0.000}_{-0.017}, \crn
&& \sin^2(2\theta_{13})=(9.3\pm0.8)\times 10^{-2},\,\,\, \Delta m^2_{21}=(7.53\pm 0.18)\times 10^{-5} \mathrm{eV}^2,\crn
&& \Delta m^2_{32}=(2.52\pm 0.07)\times
10^{-3}\mathrm{eV}^2 ,\,\, (\mathrm{Inverted \,\,\,hierarchy}).\label{PDG14I}\eea
These large neutrino mixing angles are completely different from the quark
 mixing ones defined by the Cabibbo- Kobayashi-Maskawa (CKM) matrix \cite{CKM, CKM1} , and they cannot be explained by the Standard Model. It is an interesting challenge to formulate dynamical
principles that can lead to the flavor mixing patterns for quarks
and leptons given in a completely natural way as first
approximations. This has stimulated work on flavor symmetries and non-Abelian discrete
 symmetries are considered to be the most attractive candidate to
formulate dynamical principles that can lead to the flavor
mixing patterns for quarks and lepton. There are many recent models based on
the non-Abelian discrete symmetries, such as $A_4$  (Refs.\citen{A41, A42, A43, A44, A45,A46,
A47, A48, A49, A410, A411, A412, A413, A414, A415, A416, A417, A418}) , $A_5$(Refs.\citen{A51, A52,
 A53, A54, A55, A56, A57, A58, A59, A510, A511, A512, A513}) , $S_3$ (Refs.\citen{S31,S32,S33,S34,S35,S36,S37,S38,S39,S310,S311,S312,S313,S314,S315,S316,S317,
S318,S319,S320,S321,S322,S323,S324,S325,S326,S327,S328,S329,S330,S331,S332,
S333,S334,S335,S336,S337,S338,S339,S340,S341,S342}) , $S_4$ (Refs.\citen{S41,S42,S43,S44,S45,S46,S47,S48,S49,S410,S411,S412,S413,S414,S415,S416,S417,
S418,S419,S420,S421,S422,S423,S424,S425,S426,S427,S428,S429}), $D_4$ (Refs.\citen{D41,D42,D43,D44,D45,D46,D47,D48,D49,D410,D411,D412}), $D_5$ (Refs.\citen{D51, D52}) , $T'$ (Refs.\citen{Tp1,Tp2,Tp3,Tp4,Tp5}) and so forth. In our previous works\cite{dlshA4, dlsvS4, dlnvS3, VLongD4, vlS4, vlS3, vlT7, vD4,vT7} , the discrete symmetries
have been explored to the 3-3-1 models. In Ref. \citen{dlsvS4} we have studied the 3-3-1 model with neutral fermions based
  on $S_4$ group, in which most of the Higgs multiplets are in triplets under $S_4$
    except $\chi$ lying  in a singlet, and the exact tribimaximal form \cite{hps1,hps2,hps3,hps4} is obtained,
     where $\theta_{13}= 0$.

As we know, the recent considerations have implied $\theta_{13}\neq 0$ \cite{A41, A42, A43, A44,
A45,A46, A47, A48, A49, A410, A411, A412, A413, A414, A415, A416, A417, A418, S31,S32,S33,S34,S35,S36,S37,S38,S39,S310,
S311,S312,S313,S314,S315,S316,S317,
S318,S319,S320,S321,S322,S323,S324,S325,S326,S327,S328,
S329,S330,S331,S332,S333,S334,S335,S336,S337,S338,S339,S340,S341,S342, S41,S42,S43,S44,S45,S46,S47,S48,S49,
S410,S411,S412,S413,S414,S415,S416,S417,
S418,S419,S420,S421,S422,
S423,S424,S425,S426,S427,S428,S429} , but small as given in Eqs. (\ref{PDG14N}) and (\ref{PDG14I}). This problem has been improved in Ref. \citen{dlnvS3} by adding a new triplet $\rho$ put in $\underline{1}'$
     under $S_3$ and another antisextet $s'$ put in $\underline{2}$ under $S_3$, in which $s'$ is regarded as
     a small perturbation, or a new triplet $\rho$ put in $\underline{1}''$ under $D_4$ regarded as a small perturbation \cite{VLongD4} .
     Therefore the models contain up to eight Higgs multiplets, and the scalar potential
of the model is quite complicated.

In this paper, we introduce another $SU(3)_L$ antisextet lying in $\underline{2}$ under $S_4$ which
 can result the non-zero $\theta_{13}$ without perturbation.
The rest of this work is organized
as follows. In Sec. \ref{model} we review some main results from Ref. \citen{dlsvS4}. Sec. \ref{neutrino} is devoted for the neutrino mass and
mixing. Sec. \ref{remark} presents the remark on the vacuum alignments and $\rho$ parameter. We summarize our results in the
Sec. \ref{conclus}. \ref{apa} is devoted to $S_4$ group
with its Clebsch-Gordan coefficients. \ref{apt} presents
the lepton numbers and lepton parities of model particles.  
\ref{S4breaking} provides the breakings of $S_4$ group by triplets $\underline{3}$ and $\underline{3}'$.
\section{\label{model}The model}

The fermions in this model under $[\mathrm{SU}(3)_L,
\mathrm{U}(1)_X, \mathrm{U}(1)_\mathcal{L},\underline{S}_4]$
symmetries, respectively, transform as \cite{dlsvS4}
\bea \psi_{L} &\equiv&
\psi_{1,2,3L}=\left(
  \begin{array}{c}
    \nu_{1,2,3L} \\
    l_{1,2,3L} \\
    N^c_{1,2,3R} \\
  \end{array}
\right)\sim [3,-1/3,2/3,\underline{3}],\crn
l_{1R}&\sim&[1,-1,1,\underline{1}],\hs l_R\equiv l_{2,3R}\sim[1,-1,1,\underline{2}],\crn
Q_{3L}&=& \left(
  \begin{array}{c}
    u_{3L} \\
    d_{3L} \\
    U_{L} \\
  \end{array}
\right)\sim[3,1/3,-1/3,\underline{1}],\label{fermic}\\
 Q_{L}&\equiv& Q_{1,2 L}=
\left(
  \begin{array}{c}
    d_{1,2L} \\
    -u_{1,2L} \\
    D_{1,2L} \\
  \end{array}
\right)\sim[3^*,0,1/3,\underline{2}], \crn
u_{R}&\equiv&u_{1,2,3R}\sim[1,2/3,0,\underline{3}],\hs
 d_{R}\equiv d_{1,2,3R}\sim[1,-1/3,0,\underline{3}],\crn
U_R&\sim&[1,2/3,-1,\underline{1}],\hs
D_R\equiv D_{1,2R}\sim[1,-1/3,1,\underline{2}],\nn\eea where the numbered
subscripts on field indicate respective families and define components of their $S_4$ multiplet representation. Note that the $\underline{2}$ for
quarks meets the requirement of anomaly cancelation where the last two
left-quark families are in $3^*$ while the first one as well as
the leptons are in $3$ under $SU(3)_L$. All the $\mathcal{L}$ charges of the model
multiplets are listed in the square brackets.

To generate masses for the charged leptons, we have introduced two $SU(3)_L$ scalar
triplets $\phi$ and $\phi'$ lying in $\underline{3}$ and $\underline{3}'$ under $S_4$,
respectively, with the VEVs
$\langle \phi \rangle = (v,v,v)$ and $\langle \phi' \rangle =
(v',v',v')$ written as those of $S_4$ components \cite{dlsvS4} , i.e, $S_4$ is broken
into $Z_3$ that consists of the elements\footnote{With the VEV alignment: $\langle \phi_1\rangle=\langle
\phi_2\rangle =\langle \phi_3\rangle \neq 0$,  $S_{4}$ group is  broken into $S_3$ which consisting of the elements \{$1, T, T^2, TSTS^2, STS^2, S^2TS$\}; with the VEV alignment: $\langle \phi'_1\rangle=\langle
\phi'_2\rangle =\langle \phi'_3\rangle \neq 0$, $S_{4}$ is broken into $Z_3$ that consists of the elements \{$1, T, T^2$\} as presented in \ref{S4breaking}.} $\{1, T,  T^2\}$ .
 From the invariant Yukawa interactions
for the charged leptons, we obtain
$ m_e=\sqrt{3}h_1 v ,\,\, m_\mu= \sqrt{3}(h_2 v - h_3v'),\,\,  m_\tau=\sqrt{3}(h_2 v+h_3v'),$
  and the left and right-handed charged leptons mixing matrices are
   given  \cite{dlsvS4}
    \be U_L=\fr{1}{\sqrt{3}}\left(%
\begin{array}{ccc}
  1 & 1 & 1 \\
  1 & \om & \om^2 \\
  1 & \om^2 & \om \\
\end{array}%
\right),\hs U_R=1.\label{lep}\ee

In similarity to the charged lepton sector, to generate the quark
masses, we have additionally introduced three scalar Higgs
triplets $ \chi, \,
   \eta, \, \eta'$ lying in $\underline{1}$, $\underline{3}$ and $\underline{3}'$
    under $S_4$, respectively. Quark masses can be derived from the
    invariant Yukawa interactions for quarks with supposing that the VEVs
    of $\eta$, $\eta'$ and $\chi$ are $u$, $u'$
and $v_\chi$, respectively, where $u=\langle \eta^0_1\rangle$, $u'=\langle
\eta'^0_1\rangle$, $v_\chi=\langle \chi^0_3\rangle$ and the other VEVs
$\langle \eta^0_3\rangle$, $\langle \eta'^0_3\rangle$, and
$\langle\chi^0_1\rangle$ vanish due to the lepton parity
conservation. The exotic quarks get masses $m_U=f_3 v_\chi$ and $m_{D_{1,2}}=f v_\chi$. The masses of ordinary
up-quarks and down-quarks are:
\bea
  &&m_u=-\sqrt{3}(h^u v+h'^u v') ,\, m_c = -\sqrt{3}(h^u v-h'^u v') ,\, m_t =\sqrt{3}h^u_3 u, \crn
  && m_d =\sqrt{3}(h^d u+h'^d u'),\, m_s=\sqrt{3}(h^d u-h'^d u'),\,\,\, m_b=\sqrt{3}h^d_3 v.\label{quarkmass}\eea
  The unitary matrices, which couple
the left-handed up- and down-quarks to those in the mass bases,
are $U^u_L=1$ and $U^d_L=1$, respectively. Therefore we get the
quark mixing matrix $U_\mathrm{CKM}=U^{d\dagger}_L
U^u_L=1$. For a detailed study
 on charged lepton and quark mass, the reader is referred to
 Ref. \citen{dlsvS4}. In this work, we add a new $SU(3)_L$ anti-sextet lying in $\underline{2}$ under $S_4$
 responsible for the non- zero $\theta_{13}$ without perturbation which is
  different from those in Refs. \citen{dlsvS4, dlnvS3, VLongD4} . The vacuum alignments and the gauge boson
   masses and mixings are similar to those in Refs. \citen{DongHLT, VLongD4} so we will not discuss
   it further in this work.

\section{\label{neutrino} Neutrino mass and mixing}
In this type of the models, the neutrino masses arise from the couplings of $\bar{\psi}^c_{L} \psi_{L}$
 to scalars, where $\bar{\psi}^c_{L} \psi_{L}$ transforms as $3^*\oplus 6$ under
$\mathrm{SU}(3)_L$ and $\underline{1}\oplus \underline{2}\oplus
\underline{3}\oplus \underline{3}'$ under $S_4$. For the known
scalar triplets $(\phi, \phi', \chi,\eta, \eta')$, the available
interactions are only $(\bar{\psi}^c_{L} \psi_{L})\phi$ and
$(\bar{\psi}^c_{L} \psi_{L})\phi'$, but explicitly suppressed
because of the $\mathcal{L}$-symmetry. We will therefore propose
new SU(3)$_L$ antisextets, lying in either $\underline{1}$,
$\underline{2}$, $\underline{3}$, or $\underline{3}'$ under $S_4$,
which interact with  $\bar{\psi}^c_{L}\psi_{L}$ to produce masses
for the neutrinos. In Ref.\citen{dlsvS4} we have introduced two
$SU(3)_L$ antisextets $\si, s$ transform as follows \bea \sigma&=&
\left(%
\begin{array}{ccc}
  \sigma^0_{11} & \sigma^+_{12} & \sigma^0_{13} \\
  \sigma^+_{12} & \sigma^{++}_{22} & \sigma^+_{23} \\
  \sigma^0_{13} & \sigma^+_{23} & \sigma^0_{33} \\
\end{array}%
\right)\sim [6^*,2/3,-4/3,\underline{1}], \crn
 s&=&
\left(%
\begin{array}{ccc}
  s^0_{11} & s^+_{12} & s^0_{13} \\
  s^+_{12} & s^{++}_{22} & s^+_{23} \\
  s^0_{13} & s^+_{23} & s^0_{33} \\
\end{array}%
\right)\sim [6^*,2/3,-4/3,\underline{3}],\label{s} \eea
with the VEV of $s$ is set as $(\langle s_1\rangle,0,0)$ under $S_4$, where
\be
\langle s_1\rangle=\left(%
\begin{array}{ccc}
  \la_{s} & 0 & v_{s} \\
  0 & 0 & 0 \\
  v_{s} & 0 & \Lambda_{s} \\
\end{array}%
\right),\label{vevs}\ee
and the VEV of $\sigma$ is \be
\langle \sigma \rangle=\left(%
\begin{array}{ccc}
  \la_\sigma & 0 & v_\sigma \\
  0 & 0 & 0 \\
  v_\sigma & 0 & \La_\sigma \\
\end{array}%
\right).\label{vevsi}\ee With these $SU(3)_L$ anti-sextets, the
exact tribimaximal form was obtained, in which $\theta_{13}=0$
\cite{dlsvS4} . However, the recent experimental data have implied
$\theta_{13}\neq 0$ as given in Eqs. (\ref{PDG14N}) and (\ref{PDG14I}). So that we need
to modify the neutrino mass matrix to fit the recent data.

Notice that the VEV alignment as in (\ref{vevs}), $S_4$ is broken into a group which is isomorphic
to Klein four group \cite{S410} that consists of the elements $\mathcal{K}=\{1, S^2, TSTS^2, TST\}$.
To obtain a realistic neutrino spectrum, in this work we additionally introduce another $SU(3)_L$
 anti-sextet ($s'$) which lies in $\underline{2}$ under $S_4$ and responsible for the breaking
 $\mathcal{K}\rightarrow Z_2$. This happens in any case
below:  $\langle s' \rangle = (\langle s'_1 \rangle, 0)$, with
\be
\langle s'_1\rangle=\left(%
\begin{array}{ccc}
  \la'_{s} & 0 & v'_{s} \\
  0 & 0 & 0 \\
  v'_{s} & 0 & \La'_{s} \\
\end{array}%
\right).\label{vevsp}\ee The VEV alignment of $s'$ as in
(\ref{vevsp}) will break $\mathcal{K}$ into $Z_2$ that consists of
the elements $\{1, A^2\}$ (instead of $S_4$ is broken into another
Klein four group \cite{S410} that consists of the elements $\{1,
S^2, TS^2T^2, T^2S^2T\}$).

In calculation, combining both cases we have the Yukawa
interactions responsible for neutrino mass: \bea
-\mathcal{L}_\nu&=&\fr 1 2 x (\bar{\psi}^c_L
\psi_L)_{\underline{1}}\sigma+\fr 1 2 y (\bar{\psi}^c_L
\psi_L)_{\underline{3}}s +\fr 1 2 z (\bar{\psi}^c_L
\psi_L)_{\underline{2}}s'+H.c.\crn &=&
\frac{x}{2}(\bar{\psi}^c_{1L}\psi_{1L}+\bar{\psi}^c_{2L}\psi_{2L}+\bar{\psi}^c_{3L}\psi_{3L})\sigma\crn
&+&\frac{y}{2}\left[(\bar{\psi}^c_{2L}\psi_{3L}+\bar{\psi}^c_{3L}\psi_{2L})s_1+(\bar{\psi}^c_{3L}\psi_{1L}
+\bar{\psi}^c_{1L}\psi_{3L})s_2
+(\bar{\psi}^c_{1L}\psi_{2L}+\bar{\psi}^c_{2L}\psi_{1L})s_3\right]\crn
&+&
\frac{z}{2}\left[(\bar{\psi}^c_{1L}\psi_{1L}+\om^2\bar{\psi}^c_{2L}\psi_{2L}+\om
\bar{\psi}^c_{3L}\psi_{3L})s'_2
+(\bar{\psi}^c_{1L}\psi_{1L}+\om\bar{\psi}^c_{2L}\psi_{2L}+\om^2
\bar{\psi}^c_{3L}\psi_{3L})s'_1\right]\crn &+&H.c.\label{yn}\eea
The mass Lagrangian for the neutrinos is given by \bea
-\mathcal{L}^{\mathrm{mass}}_\nu &=&\fr 1 2 x
\left(\la_{\si}\bar{\nu}^c_{1 L}\nu_{1L}+v_{\si}\bar{\nu}^c_{1
L}N^c_{1R}+ v_{\si}\bar{N}_{1R}\nu_{1L}
+\La_{\si}\bar{N}_{1R}N^c_{1R}\right.\crn
&+&\left.\la_{\si}\bar{\nu}^c_{2 L}\nu_{2L}+v_{\si}\bar{\nu}^c_{2
L}N^c_{2R}+v_{\si}\bar{N}_{2R}\nu_{2L}+\La_{\si}\bar{N}_{2R}N^c_{2R}\right.\crn
&+&\left.\la_{\si}\bar{\nu}^c_{3 L}\nu_{3L} +v_{\si}\bar{\nu}^c_{3
L}N^c_{3R}+v_{\si}\bar{N}_{3R}\nu_{3L}+\La_{\si}\bar{N}_{3R}N^c_{3R}\right)\crn
&+& \frac{y}{2}\left[\la_{s}(\bar{\nu}^c_{2
L}\nu_{3L}+\bar{\nu}^c_{3
L}\nu_{2L})+v_{s}\left(\bar{\nu}^c_{2L}N^c_{3R}+\bar{\nu}^c_{3L}N^c_{2R}\right)\right.\crn
&+&\left.v_{s}(\bar{N}_{2R}\nu_{3L}+\bar{N}_{3R}\nu_{2L})+\La_{s}(\bar{N}_{2R}N^c_{3R}
+\bar{N}_{2R}N^c_{3R})\right]\crn
&+& \fr z 2 \left[\left(\la'_s\bar{\nu}^c_{1
L}\nu_{1L}+v'_s\bar{\nu}^c_{1
L}N^c_{1R}+v'_s\bar{N}_{1R}\nu_{1L}+\La'_s\bar{N}_{1R}N^c_{1R}\right)\right.\crn
&+&\left. \om (\la'_s\bar{\nu}^c_{2 L}\nu_{2L} +v'_s\bar{\nu}^c_{2
L}N^c_{2R}+v'_s\bar{N}_{2R}\nu_{2L}+\La'_s\bar{N}_{2R}N^c_{2R})\right.\crn
&+&\left.\om^2 \left(\la'_s\bar{\nu}^c_{3
L}\nu_{3L}+v'_s\bar{\nu}^c_{3 L}N^c_{3R}+
v'_s\bar{N}_{3R}\nu_{3L}+\La'_s\bar{N}_{3R}N^c_{3R}\right)\right]
+H.c. \label{S4Lanu}\eea We can rewrite the mass Lagrangian for
the neutrinos in the matrix form: \bea
-\mathcal{L}^{\mathrm{mass}}_\nu&=&\fr 1 2 \bar{\chi}^c_L M_\nu
\chi_L+ H.c.,\crn \chi_L&\equiv&
\left(%
\begin{array}{c}
  \nu_L \\
  N^c_R \\
\end{array}%
\right),\hs M_\nu\equiv\left(%
\begin{array}{cc}
  M_L & M^T_D \\
  M_D & M_R \\
\end{array}%
\right),\label{nm}\eea where $\nu=(\nu_{1},\nu_{2},\nu_{3})^T$ and
$N=(N_1,N_2,N_3)^T$. The mass matrices are then obtained by
\[ M_{L,D,R}=\left(%
\begin{array}{ccc}
  a_{L,D,R}+d_{L,D,R} & 0 & 0 \\
  0 & a_{L,D,R}+\om d_{L,D,R} & b_{L,D,R} \\
  0 & b_{L,D,R} & a_{L,D,R}+\om^2 d_{L,D,R} \\
\end{array}%
\right),\] where \bea a_{L} & =&\la_{\si}x,\,\,\,\,\,\,\,
a_{D}=v_{\si}x,\hs a_{R} =\La_{\si}x,\crn
 b_{L} & =&\la_s y,\hs b_{D}=v_{s}y,\hs b_{R} =\La_{s}y, \label{abcLDR}\\
 d_{L} & = &\la'_{s}z,\hs d_{D}=v'_{s}z,\hs d_{R}=\La'_{s}z.\nn \eea

The VEVs $\La_{\sigma,s}$ break the 3-3-1 gauge symmetry down to
that of the standard model, and provide the masses for the neutral
fermions $N_R$ and the new gauge bosons: the neutral $Z'$ and the
charged $Y^{\pm}$ and $X^{0,0*}$. The $\la_{\sigma,s}$ and
$v_{\sigma,s}$ belong to the second stage of the symmetry breaking
from the standard model down to the $\mathrm{SU}(3)_C \otimes
\mathrm{U}(1)_Q$ symmetry, and contribute the masses to the
neutrinos. Hence, to keep a consistency we assume that
$\La_{\sigma,s}\gg v_{\sigma,s},\la_{\sigma,s}$ \cite{dlsvS4} . The
natural smallness of the lepton number violating VEVs
$\la_{\sigma,s}$ and $v_{\sigma,s}$ was explained in Ref.\citen{dlsvS4} . Three active-neutrinos therefore gain masses via a
combination of type I and type II seesaw mechanisms derived from
(\ref{nm}) as \be M_{\mathrm{eff}}=M_L-M_D^TM_R^{-1}M_D=
\left(%
\begin{array}{ccc}
  A & 0 & 0 \\
  0 & B_1 & D \\
  0 & D & B_2 \\
\end{array}%
\right),\label{neu}\ee where \bea
A&=&a_L+d_L-\fr{(a_D+d_D)^2}{a_R+d_R},\hs D=\frac{a_2-
b_2}{a^2_R+d^2_R-a_Rd_R-b^2_R}, \crn B_1&=&-\frac{a_1+ b_1 \om^2+
c_1 \om}{a^2_R+d^2_R-a_Rd_R-b^2_R},\hs B_2= \frac{a_1+ b_1 \om+
c_1 \om^2}{a^2_R+d^2_R-a_Rd_R-b^2_R},\label{AB12D}\eea with \bea
a_1&=&a^2_D a_R+2a_D(d_Dd_R-b_Db_R)+a_R(b^2_D-d_Ld_R)
-a_L(a^2_R-b^2_R+d^2_R)+a_La_Rd_R,\crn
b_1&=&a^2_Dd_R+a_R(d^2_D-d_Ld_R), \crn
c_1&=&2d_D(a_Da_R-b_Db_R)+(b^2_D+d^2_D)d_R
-d_L(a^2_R-b^2_R+d^2_R),  \label{abc12}\\
a_2&=&a^2_D b_R-2b_D(a_Da_R+d_Dd_R)+a^2_Rb_L
+b_R(b^2_D-b_Lb_R+d^2_D)+b_Ld^2_R,\crn
b_2&=&-a_Rb_Dd_D+a_Db_Rd_D-a_Db_Dd_R+a_Rb_Ld_R.\nn \eea We can
diagonalize the mass matrix (\ref{neu}) as follows: \[ U^T_\nu
M_{\mathrm{eff}} U_\nu=\mathrm{diag}(m_1,\,\, m_2,\,\, m_3), \]
where \bea m_1 &=&\fr 1 2 \left(B_1 + B_2+ \sqrt{4 D^2 +
(B_1-B_2)^2}\right),\crn
m_2&=&A,\label{m123}\\
m_3&=&\fr 1 2 \left(B_1 + B_2 -\sqrt{4 D^2 + (B_1-B_2)^2}\right),\nn\eea
and the corresponding eigenstates put in the
lepton mixing matrix:
\be U_\nu=\left(%
\begin{array}{ccc}
  0 & 1 & 0 \\
 \fr{1}{\sqrt{K^2+1}}& 0 &\fr{K}{\sqrt{K^2+1}}   \\
 -\fr{K}{\sqrt{K^2+1}} & 0 &  \fr{1}{\sqrt{K^2+1}} \\
\end{array}%
\right)\left(%
\begin{array}{ccc}
 1 & 0 & 0 \\
0 & 1 & 0 \\
0 & 0 & -i \\
\end{array}%
\right),\label{neu1}\ee where \be K =\frac{B_1 - B_2
-\sqrt{ 4 D^2+(B_1-B_2)^2}}{2D}.\label{K} \ee 
The lepton mixing
matrix is defined as  \be U_{lep}\equiv U^\dagger_L
U_\nu=\fr{1}{\sqrt{3}}\left(%
\begin{array}{ccc}
  \fr{1-K}{\sqrt{K^2+1}} & 1 &  \fr{1+K}{\sqrt{K^2+1}} \\
  \fr{\om(\om-K)}{\sqrt{K^2+1}} & 1 &  \fr{\om(K\om+1)}{\sqrt{K^2+1}} \\
  \fr{\om(1-K\om)}{\sqrt{K^2+1}} & 1 & \fr{\om(\om+K)}{\sqrt{K^2+1}}  \\
\end{array}%
\right).\left(%
\begin{array}{ccc}
 1 & 0 & 0 \\
0 & 1 & 0 \\
0 & 0 & -i \\
\end{array}%
\right).\label{Ulep}\ee
It is easily to check that $U_L$ in (\ref{lep}) is an unitary matrix. So, if  $U_\nu$ in (\ref{neu1}) is unitary then $U_{lep}$ in (\ref{Ulep}) is unitary. Here, we will only consider real values for $K$ since the unitary condition of $U_{lep}$. Furthermore, it is worth noting that in the case of the
subgroup $\mathcal{K}$ is unbroken, i.e, without contribution of
$s'$ (or $\la'_s=v'_s=\La'_s=0$), the lepton mixing matrix
(\ref{Ulep})  being equal to $U_{HPS}$ as given in (\ref{Uhps}).

The value of the Jarlskog invariant $J_{CP}$, which gives a convention-independent measure of CP violation, is defined from (\ref{Ulep}) as 
\bea
J_{CP} = \mathrm{Im}[U_{21} U_{31}^* U_{22}^* U_{32}]=\frac{1-K^2}{6\sqrt{3}(1+K^2)}. \label{J1}\eea

Until now the values of neutrino masses (or the
absolute neutrino masses) as well as the mass ordering of
neutrinos are unknown. 
The neutrino mass spectrum can be the normal hierarchy ($
|m_1|\simeq |m_2| < |m_3|$), the inverted hierarchy ($|m_3|<
|m_1|\simeq |m_2|$)
 or nearly degenerate ($|m_1|\simeq |m_2|\simeq |m_3| $). 
An upper bound on the absolute value of
neutrino  mass was found from the analysis of the cosmological
data \cite{Tegmark} 
\be m_i\leq 0.6\, \mathrm{eV},\label{upb} \ee
while the upper limit on the sum of neutrino masses given in \cite{Ade}
\be \sum^{3}_{i=1}m_i< 0.23\, \mathrm{eV} \label{upbsum}\ee
 In the case of 3-neutrino mixing, the two possible signs of $\Delta
m^2_{23}$ corresponding to two types of neutrino mass spectrum can
be provided as follows:
\ben \item[$\circ$] Normal hierarchy (NH): $
|m_1|\simeq |m_2| < |m_3|,\,\, \Delta m^2_{32}=m^2_3-m^2_2>0.$
\item [$\circ$] Inverted hierarchy (IH): $|m_3|< |m_1|\simeq |m_2|,\,\, \Delta m^2_{32} =m^2_3-m^2_2<0$.\een
As will be discussed below, the model under consideration can provide both normal and inverted
 mass hierarchy.

\subsection{Normal case ($\Delta m^2_{32}> 0$)}
In the Normal Hierarchy, combining (\ref{J1}) with the data in Ref. \citen{PDG2014}, $J_{CP}= -0.032$,
we get \bea
K&=&-1.41297,\label{Kn}
\eea 
and the lepton mixing matrices are obtained as
\bea 
U_{lep}&=&\left(%
\begin{array}{ccc}
 0.805             & \frac{1}{\sqrt{3}} &0.138\\
-0.402 + 0.119i&\frac{1}{\sqrt{3}}&0.069 + 0.697i \\
-0.402 - 0.119i& \frac{1}{\sqrt{3}} &0.069 - 0.697i\\
\end{array}%
\right)\times P,\label{Ulep11}\eea
\textbf{or}
\bea
\left|U_{lep}\right|&=&\left(%
\begin{array}{ccc}
 0.805&\hs 0.577 &\hs 0.138\\
0.420 &\hs 0.577 &\hs 0.700\\
0.420&\hs 0.577 &\hs 0.700\\
\end{array}%
\right), \label{Ulepab1}
\eea
In the standard parametrization, the lepton mixing
 matrix can be parametrized as
 \bea
 U_{PMNS} = \left(%
\begin{array}{ccc}
    c_{12} c_{13}     & s_{12} c_{13}                    & s_{13} e^{-i\delta}\\
    -s_{12} c_{23}-c_{12} s_{23} s_{13} e^{i \delta} & c_{12} c_{23}-s_{12} s_{23} s_{13} e^{i \delta} &s_{23} c_{13}\\
    s_{12} s_{23}-c_{12} c_{23} s_{13}e^{i \delta}&-c_{12} s_{23}-s_{12} c_{23} s_{13} e^{i \delta}  & c_{23} c_{13} \\
    \end{array}%
\right). \mathcal{P}, \label{Ulepg1}\eea
where $\mathcal{P}=\mathrm{diag}(1, e^{i \alpha}, e^{i \beta})$, and
$c_{ij}=\cos \theta_{ij}$, $s_{ij}=\sin \theta_{ij}$ with
$\theta_{12}$, $\theta_{23}$ and $\theta_{13}$ being the
solar, atmospheric and  reactor angles, respectively.
$\delta= [0, 2\pi]$ is the Dirac CP violation phase while $\alpha$ and
$\beta$ are two Majorana CP violation phases.
Using the parametrization in Eq. (\ref{Ulepg1}) we get
\be
J_{CP}=\frac{1}{8}\cos\theta_{13}\sin2\theta_{12}\sin2\theta_{23}\sin2
\theta_{13}\sin\delta.\label{Jp1}
\ee
With the help of (\ref{PDG14N}), (\ref{Kn}) and (\ref{Jp1}) we have $\sin\delta_{CP}=-0.9242 $, i.e, $ \delta_{CP}=-67.55^\circ$ or $ \delta_{CP}=292.45^\circ $.

From Eqs. (\ref{K}) and (\ref{Kn}) we get
\bea
B_1 &=& B_2 - 0.705241 D. \label{BDn}
\eea 
In the normal case, i.e, $\Delta m^2_{32}=m^2_{3}-m^2_{2}> 0$, taking the central values of neutrino mass squared difference from the data 
in \citen{PDG2014} as shown in (\ref{PDG14N}): $\Delta m^2_{21}=7.53\times 10^{-5}\
\mathrm{eV}^2$ and $\Delta m^2_{32}=2.44\times 10^{-3}\
\mathrm{eV}^2$, with $m_{1,2,3}$ given in Eq. (\ref{m123}), we get a solution\footnote{In fact, this system of equations has four solutions, however, these equations differ only by the sign of $m_{1,2, 3}$ that it is not appear in the neutrino oscillation experiments. So, here we only consider in
detail the solution in (\ref{B2Dn}).} (in [eV]) 
\bea
B_2 &=&-0.5\sqrt{4A^2-0.0003} -0.707729D,\crn
D&=&0.471543\left(\sqrt{A^2+2.44\times 10^{-3}}-\sqrt{A^2-7.53\times 10^{-5}}\right). \label{B2Dn}
\eea 
With $B_{1,2}$ and $D$ in Eqs. (\ref{B2Dn}) and (\ref{BDn}), $m_{1,2, 3}$ depends only on one parameter $A$, so we will consider $m_{1,2,3}$ as
functions of $A$. By using the upper bound on the absolute value of
neutrino  mass in (\ref{upb}) we can
 restrict the values of $A$:  $\left|A\right| \leq 0.6\,\mathrm{eV}$. However, in this case, $A \in (0.0087, 0.05)\, \mathrm{eV}$ or $A \in (-0.05, -0.0087)\, \mathrm{eV}$ are good regions of $A$
that can reach the realistic neutrino mass hierarchy.

\begin{figure}[ht]
\bc
\includegraphics[width=12.0cm, height=6.0cm]{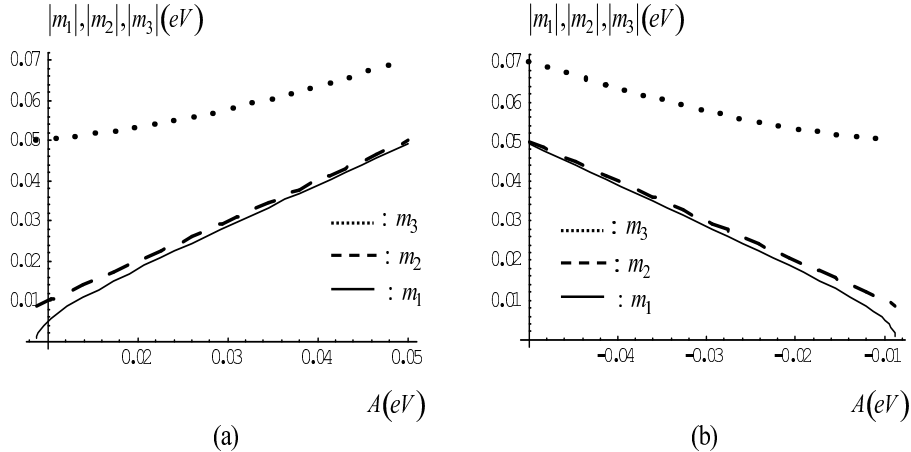}
\vspace*{-0.1cm} \caption[$|m_{1,2,3}|$ as functions of $A$ in
 the case of $\Delta m^2_{32}> 0$ with
 a) $A\in(0.00867, 0.05) \, \mathrm{eV}$ and b) $A\in(-0.05, -0.00867) \, \mathrm{eV}$.]{$|m_{1,2,3}|$ as functions of $A$ in
 the case of $\Delta m^2_{32}> 0$ with
 a) $A\in(0.00867, 0.05) \, \mathrm{eV}$ and b) $A\in(-0.05, -0.00867) \, \mathrm{eV}$.}\label{m123N}
\ec
\end{figure}

\begin{figure}[ht]
\bc
\includegraphics[width=6.0cm, height=4.5cm]{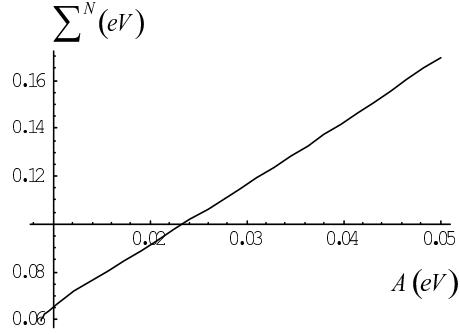}
\vspace*{-0.1cm} \caption[$\sum^N$ as a function of $A$ with
 $A\in(0.00867, 0.05) \, \mathrm{eV}$ in
 the case of $\Delta m^2_{32}> 0$.]{$\sum^N$ as a function of $A$ with
 $A\in(0.00867, 0.05) \, \mathrm{eV}$ in
 the case of $\Delta m^2_{32}> 0$.}\label{m123Ns}
\ec
\end{figure}

In Fig. \ref{m123N}, we have plotted the  absolute value  $|m_{1,2,3}|$
as functions of $A$ with  $A \in (0.0087, 0.05)\, \mathrm{eV}$ and $A \in (-0.05,-0.0087)\, \mathrm{eV}$, respectively. This figure shows
  that there exist allowed regions for values of $A$ where either normal
  or quasi-degenerate neutrino masses spectrum is achieved.
  The quasi-degenerate mass hierarchy is obtained if $\left|A\right|\in [0.05\,\mathrm{eV} , +\infty$). However, $\left|A\right|$ must be small enough because of the scale of $\left|m_{1,2,3}\right|$). The normal
  mass hierarchy will be obtained if $A$ takes the values around $(0.0087, 0.05)\,
   \mathrm{eV}$ or $(-0.05, -0.0087)\, \mathrm{eV}$. 
The sum of neutrino masses in the normal case
   $\sum^N=\sum^3_{i=1}|m_i|$ with $A \in (0.0087, 0.05)\,\mathrm{eV}$ is depicted in Fig. \ref{m123Ns} which is consistent with the upper limit given in Eq.(\ref{upbsum}).

\begin{figure}[h]
\begin{center}
\includegraphics[width=7.0cm, height=6.0cm]{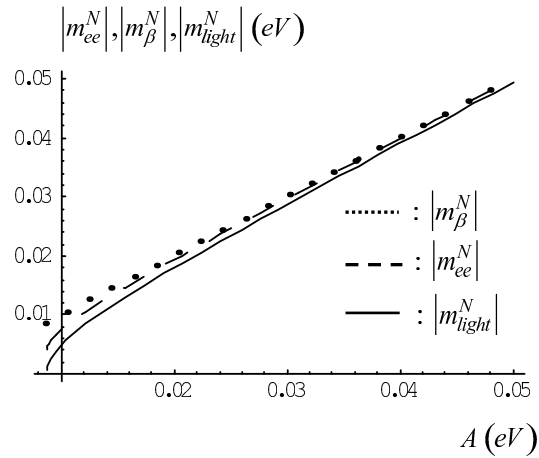}
\caption[ $|m^N_{ee}|$, $|m^N_{\beta}|$ and $|m^N_{light}| $ as functions of $A$ with $A \in (0.0087, 0.05)\,\mathrm{eV}$ in the case of $\Delta m^2_{32}> 0$]{$|m^N_{ee}|$, $|m^N_{\beta}|$ and $|m^N_{light}| $ as functions of $A$ with $A \in (0.0087, 0.05)\,\mathrm{eV}$ in the case of $\Delta m^2_{32}> 0$.}\label{meeNv}
\end{center}
\end{figure}

From the expressions (\ref{K}), (\ref{Ulep}), (\ref{BDn}) and (\ref{B2Dn}), it is easily
 to obtain the effective masses governing neutrinoless double beta decay
 \cite{betdecay1, betdecay2,betdecay3,betdecay5,betdecay6, betdecay7} , 
\bea
m^N_{ee} = \sum^3_{i=1} U_{ei}^2 \left|m_i\right| , \hs 
m^N_\beta = \left(\sum^3_{i=1} \left|U_{ei}\right|^2 m_i^2\right)^{1/2} \label{meemb} \eea
 which is plotted in Fig. \ref{meeNv} with $A \in (0.0087, 0.05)\,\mathrm{eV}$ in the case of $\Delta m^2_{32}> 0$.
We also note
that in the normal spectrum, $|m_1|\approx |m_2|<|m_3|$, so $m_1$
given in (\ref{m123}) is the lightest neutrino mass, which is denoted as $m_{1}\equiv m^N_{light}$.

To get explicit values of the model
parameters, we set $A =10^{-2}\, \mathrm{eV}$, which is safely small.
 The other physical neutrino masses are explicitly
 given as
 \bea
 |m_1|\simeq 4.97 \times10^{-3}\, \mathrm{eV},\hs |m_2|=10^{-2}\, \mathrm{eV},\hs |m_3|\simeq 5.04 \times 10^{-2} \, \mathrm{eV}.\eea
 It follows that
\bea
 && |m_{ee}^N|\simeq 7.50\times10^{-3}\, \mathrm{eV},\hs |m_{\beta}^N|=9.87\times10^{-3}\, \mathrm{eV},\label{meeN}\\
&&B_1=-3.523\times 10^{-2} \,\mathrm{eV},\,\, B_2=-2.013\times 10^{-2} \,\mathrm{eV},\,\, 
D=2.142\times 10^{-2} \,\mathrm{eV}.\eea
 This solution means a normal mass spectrum as mentioned above.
Furthermore, by assuming that\footnote{The values of the parameters $\la_s, \la'_s, \la_\si, v_s, v'_s, v_\si, \La_s, \La'_s, \La_\si$ have not been confirmed by experiment, however, their hierarchies were given in Ref. \citen{dlnvS3}. The parametres in Eqs. (\ref{assum}) and (\ref{xyza}) is a set of the model parameters that can fit the experimental data on neutrino given in (\ref{PDG14N}).}
\bea
\la_s=\la'_{s}=\la_{\si}=1\,\mathrm{eV},\,\, v_s=v'_s=v_\si,\,  \La'_s= \La_\si=\La_s, \La_s=a v^2_s,\label{assum}\eea
 we obtain a solution
 \bea
 x&\simeq&(2.0+ 0.2i)\times 10^{-3}, \,\, 
 y\simeq -(6.1+ 0.61i)\times 10^{-3},\crn
 z &=&-(4.85 + 0.48)\times 10^{-3}, \,\,  a\simeq 0.222 + 0.017 i. \label{xyza}
 \eea
\subsection{Inverted case ($\Delta m^2_{32}< 0$)}

For inverted hierarchy, the data in Ref. \citen{PDG2014} implies $J_{CP}= -0.029$. Hence, we get \bea
K&=&-1.36483,\label{Ki}
\eea 
and the lepton mixing matrices are obtained as
\bea 
U_{lep}&=&\left(%
\begin{array}{ccc}
 0.807             & \frac{1}{\sqrt{3}} &-0.125\\
-0.403 + 0.108i&\frac{1}{\sqrt{3}}&0.062 + 0.699i \\
-0.403 - 0.108i& \frac{1}{\sqrt{3}} &0.062 - 0.699i\\
\end{array}%
\right)\times P,\label{Ulep12}\eea
\textbf{or}
\bea
\left|U_{lep}\right|&=&\left(%
\begin{array}{ccc}
 0.807&\hs 0.577 &\hs 0.125\\
0.418 &\hs 0.577 &\hs 0.701\\
0.418&\hs 0.577 &\hs 0.701\\
\end{array}%
\right). \label{Ulepab2}
\eea
Combining (\ref{PDG14I}), (\ref{Jp1}) and (\ref{Ki}) yields $\sin\delta_{CP}=-0.8371 $, i.e, $ \delta_{CP}=-56.84^\circ$ or $ \delta_{CP}=303.14^\circ $.

From Eqs. (\ref{K}) and (\ref{Ki}) we get
\bea
B_1 &=& B_2 - 0.632138 D. \label{BDi}
\eea 
In the inverted case, $\Delta m^2_{32}=m^2_{3}-m^2_{2}< 0$, taking the central values of neutrino mass squared difference from the data 
in Ref. \citen{PDG2014} as shown in (\ref{PDG14I}): $\Delta m^2_{21}=7.53\times 10^{-5}\
\mathrm{eV}^2$ and $\Delta m^2_{32}=-2.52\times 10^{-3}\
\mathrm{eV}^2$, with $m_{1,2,3}$ given in Eq. (\ref{m123}), we get a solution\footnote{Similarly to the normal case, there are four solutions in the inverted hierarchy. Here we only consider in
detail the solution in (\ref{B2Di}).} (in [eV]) 
\bea
B_2 &=&-0.5\sqrt{4A^2-0.0003} -0.732692 D,\crn
D&=&-0.476753\left(\sqrt{A^2+2.52\times 10^{-3}}+\sqrt{A^2-7.53\times 10^{-5}}\right). \label{B2Di}
\eea 
With $B_{1,2}$ and $D$ in Eqs. (\ref{BDi}) and (\ref{B2Di}), $m_{1,2, 3}$ depends only on one parameter $A$, so we will consider $m_{1,2,3}$ as
functions of $A$. In this case, $A \in (0.05, 0.1)\, \mathrm{eV}$ or $A \in (-0.1, -0.05)\, \mathrm{eV}$ are good regions of $A$
that can reach the realistic neutrino mass hierarchy. 

\begin{figure}[ht]
\bc
\includegraphics[width=12.0cm, height=6.0cm]{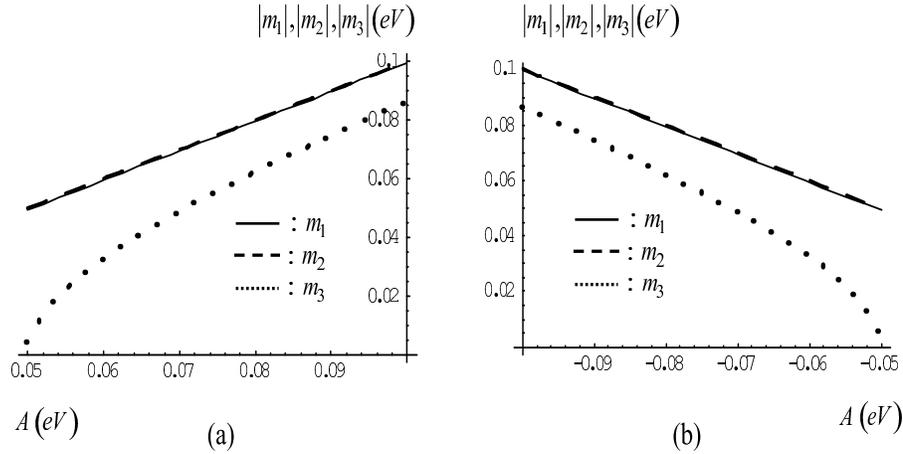}
\vspace*{-0.1cm} \caption[$|m_{1,2,3}|$ as functions of $A$ in
 the case of $\Delta m^2_{32}< 0$ with
 a) $A\in(0.05, 0.1) \, \mathrm{eV}$ and b) $A\in(-0.1, -0.05) \, \mathrm{eV}$.]{$|m_{1,2,3}|$ as functions of $A$ in
 the case of $\Delta m^2_{32}< 0$ with
 a) $A\in(0.05, 0.1) \, \mathrm{eV}$ and b) $A\in(-0.1, -0.05) \, \mathrm{eV}$.}\label{m123I}
\ec
\end{figure}

In Fig. \ref{m123I}, we have plotted the  absolute value $|m_{1,2,3}|$
as functions of $A$ with  $A \in (0.05, 0.1)\, \mathrm{eV}$ and $A \in (-0.1, -0.05)\, \mathrm{eV}$, respectively. We see that there exist allowed regions for values of $A$ where either inverted or quasi-degenerate neutrino masses spectrum is achieved.
  The quasi-degenerate mass hierarchy is obtained if $\left|A\right|\in [0.1\,\mathrm{eV}, +\infty$). However, $\left|A\right|$ must be small enough because of the scale of $\left|m_{1,2,3}\right|$. The inverted mass hierarchy will be obtained if $\left|A\right|$ takes the values around $(0.05, 0.1)\,
   \mathrm{eV}$. 
The sum of neutrino masses in the normal case
   $\sum^N=\sum^3_{i=1}|m_i|$ with $A \in (0.0087, 0.05)\,\mathrm{eV}$ is depicted in Fig. \ref{m123Ns} which is consistent with the upper limit given in Eq.(\ref{upbsum}).
\begin{figure}[ht]
\bc
\includegraphics[width=6.0cm, height=4.5cm]{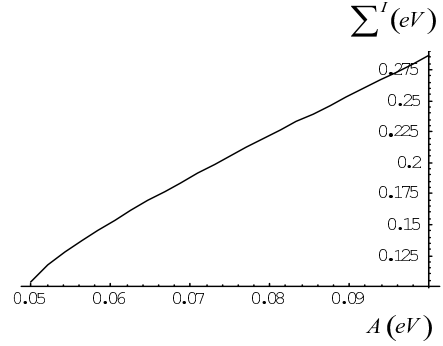}
\vspace*{-0.1cm} \caption[$\sum^I$ as a function of $A$ with
 $A\in(0.05, 0.1) \, \mathrm{eV}$ in
 the case of $\Delta m^2_{32}< 0$.]{$\sum^I$ as a function of $A$ with
 $A\in(0.05, 0.1) \, \mathrm{eV}$ in
 the case of $\Delta m^2_{32}< 0$.}\label{m123Is}
\ec
\end{figure}
The effective masses governing neutrinoless double beta decay defined in (\ref{meemb}) is plotted in Fig. \ref{meeI} with $A \in (0.05, 0.1)\,\mathrm{eV}$ in the case of $\Delta m^2_{32}< 0$.
We also note
that in the inverted spectrum, $|m_3|\approx |m_2|\simeq|m_1|$, so $m_3$
given in (\ref{m123}) is the lightest neutrino mass, which is denoted as $m_{3}\equiv m^I_{light}$.
\begin{figure}[h]
\begin{center}
\includegraphics[width=7.0cm, height=6.0cm]{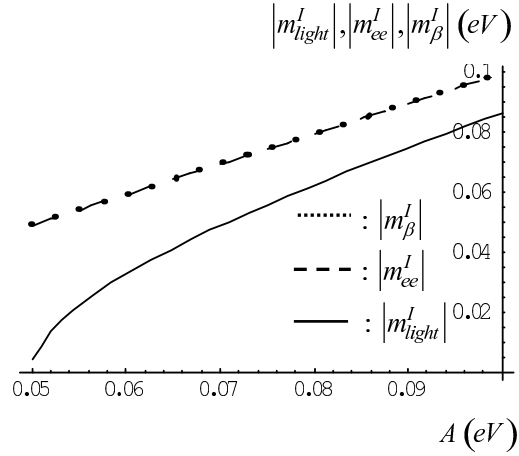}
\caption[ $|m^I_{ee}|$, $|m^I_{\beta}|$ and $|m^I_{light}| $ as functions of $A$ with $A \in (0.05, 0.1)\,\mathrm{eV}$ in the case of $\Delta m^2_{32}< 0$]{$|m^I_{ee}|$, $|m^I_{\beta}|$ and $|m^I_{light}| $ as functions of $A$ with $A \in (0.05, 0.1)\,\mathrm{eV}$ in the case of $\Delta m^2_{32}< 0$.}\label{meeI}
\end{center}
\end{figure}

To get explicit values of the model
parameters, we set $A =5\times10^{-2}\, \mathrm{eV}$, which is safely small.
 The other physical neutrino masses are explicitly
 given as
 \bea
 |m_1|\simeq 4.924 \times10^{-2}\, \mathrm{eV},\hs |m_2|=5\times10^{-2}\, \mathrm{eV},\hs |m_3|\simeq 4.472 \times 10^{-3} \, \mathrm{eV}.\eea
 It follows that
\bea
 && |m_{ee}^I|\simeq 4.88\times10^{-2}\, \mathrm{eV},\hs |m_{\beta}^I|=4.91\times10^{-2}\, \mathrm{eV},\label{meeI}\\
&&B_1=(1.72 - 0.29i)\times 10^{-2} \,\mathrm{eV},\hs B_2=(3.204 - 0.156i)\times 10^{-2} \,\mathrm{eV},\crn
&& D=(2.348 + 0.213i)\times 10^{-2} \,\mathrm{eV}.\eea
 This solution means an inverted mass spectrum. Furthermore, by assuming that \footnote{The values of the parameters $\la_s, \la'_s, \la_\si, v_s, v'_s, v_\si, \La_s, \La'_s, \La_\si$ have not been confirmed by experiment, however, their hierarchies were given in Ref. \citen{dlnvS3}. The parametres in Eqs. (\ref{assumi}) and (\ref{xyzai}) is a set of the model parameters that can fit the experimenta data on neutrino given in (\ref{PDG14I}).}
\bea
&&\la_s=\la'_{s}=\la_{\si}=a ,\,\, v_s=v'_s=-v_\si,\,  \La'_s= \La_s=-\La_\si, \crn
&&\La_s=v^2_s,\,\, \La'_s=v'^2_s,\,\, \La_\si=-v^2_\si, \label{assumi}\eea
 we obtain a solution 
 \bea
 x&\simeq&(3.192- 0.452i)\times 10^{-2}, \,\,\,\,\, 
 y\simeq (-2.563 + 0.294i)\times 10^{-2},\crn
 z &=&-(1.910+ 0.825i)\times 10^{-2}, \,\,  a\simeq 0.105 - 0.186i. \label{xyzai}
 \eea

\section{\label{remark} Remark on the vacuum alignments and $\rho$ parameter}
In the model under consideration, to generate masses for all fermions, we need eight Higgs scalars $\phi, \phi', \chi, \eta, \eta',\si, s, s'$. It is important to note that $\chi$ and $s'$ do not break $S_4$ since they are put in $\underline{1}$ under $S_4$ while $s', \phi, \eta; \phi', \eta'$ can break $S_4$ into its subgroups since they are put in non-trivial representations $\underline{2}, \underline{3}, \underline{3}'$ of $S_4$. The breaking of $S_4$ group depends on the vacuum alignment of the flavons. 

For doublets $\underline{2}$ $(s')$ we have two followings alignments. The first alignment, $0\neq\langle s'_1\rangle\neq\langle s'_2\rangle=0$ or $0\neq\langle s'_2\rangle\neq\langle s'_1\rangle=0$  or $0\neq\langle s'_1\rangle\neq\langle s'_2\rangle\neq 0$ then $S_{4}$ is broken into a group which is isomorphic to Klein four group \cite{KG} that consists of the elements $\{1, TS^2T^2, S^2, T^2S^2T\}$. The second alignment, $\langle s'_1\rangle =\langle s'_2\rangle\neq0$ then $S_{4}$ is broken into $D_{4}$ consists of the elements $\{1, TSTS^2, TST, S, S^3, TS^2T^2, S^2, T^2S^2T\}$.
For triplets $\underline{3}$ and $\underline{3}'$ the breakings of $S_4$ are given in \ref{S4breaking}.

To obtain a realistic neutrino spectrum, in this work, we
argue that the breaking $S_4\rightarrow Z_3$ is taken place in charged lepton sector while both breakings $S_4\rightarrow \mathcal{K}$ and
$\mathcal{K}\rightarrow Z_2$ must be taken place in neutrino sector.
 
Note that $\La_\si, \La_s, \La_{\si'}$ are needed to the same
order and not to be so large that can naturally be taken at TeV
scale as the VEV $v_\chi$ of $\chi$. This is because $v_\si, v_s$
and $v_{\si'}$ carry lepton number, simultaneously breaking the
lepton parity which is naturally constrained to be much smaller
than the electroweak scale \cite{dlshA4, dlsvS4, dlnvS3,e331v1,
e331v2} . This is also behind a theoretical fact that $v_\chi$,
$\La_\si, \La_s, \La_{\si'}$ are scales for the gauge symmetry breaking in the first
stage from $\mathrm{SU}(3)_L\otimes \mathrm{U}(1)_X\rightarrow
\mathrm{SU}(2)_L\otimes \mathrm{U}(1)_Y$ in the original form of
3-3-1 models \cite{e331v1, e331v2, clong, DongLongdh} . They provide masses
for the new gauge bosons $Z'$, $X$ and $Y$. Also, the exotic
quarks gain masses from $v_\chi$ while the neutral fermions masses
arise from $\La_\si, \La_s, \La_{\si'}$. The second stage of the
gauge symmetry breaking  from $\mathrm{SU}(2)_L\otimes
\mathrm{U}(1)_Y\rightarrow \mathrm{U}(1)_Q$ is achieved by the
electroweak scale VEVs such as $u, v$ responsible for ordinary
quark masses. In combination with those of type II seesaw as
determined, in this type of the model, the following limit is often taken into
account  \cite{dlshA4, dlsvS4, dlnvS3, e331v1, e331v2, clong, DongLongdh} : \bea
(\mathrm{eV})^2 \sim \la^2_\si,\la^2_s, \la^2_{\si'}&\ll& v^2_\si,
v^2_s, v^2_{\si'}\ll u^2, u'^2, v^2, v'^2\crn
&\ll& v_\chi^2\sim
\La^2_\si\sim\La^2_s\sim \La^2_{\si'}\sim(\mathrm{TeV})^2.
\label{limmitv}\eea

On the other hand, our model can modify the precision electroweak parameter such as $\rho$ parameter at the tree-level.
To see this let us approximate the masses of $W$ and $Z$ bosons 
\footnote{We have used the notation $s_W=\sin\theta_W,\,
c_W=\cos\theta_W,\, t_W=\tan\theta_W$, and the continuation of the
gauge coupling constant $g$ of the $\mathrm{SU}(3)_L$ at the
spontaneous symmetry breaking point \cite{DongHLT, DongLongdh,VLongD4}  $t=\frac{3\sqrt{2}s_W}{\sqrt{3-4s^2_W}}$ was used.}: \bea
M^2_W &\simeq& 2g^2(3u^2-v^2_{\si}),\hs M^2_Z \simeq \frac{g^2u^2}{c^2_W}\left(6-\frac{v^2_\si}{12}\right),\label{MWZ}\\
M^2_Y&\simeq&\frac{g^2}{2}\left(6\La^2_\si+4\La^2_{\si'}+2\La^2_{\si'}+v_\chi^2\right).\label{MY}\eea
The $\rho$ parameter is defined as \bea \hs \rho=\fr{M^2_W}{c^2_W M^2_Z}\simeq
1-\fr{v^2_s}{3u^2}. \label{rho}\eea 
It is easily to see that the $\rho$ parameter in
(\ref{rho}) is absolutely close to the unity since $v^2_s\ll u^2$ and this is in agreement with
the data in Ref.\citen{PDG2014}.

The mixings between the charged gauge bosons $W-Y$ and the neutral ones $Z'-W_4$ are in the same
order since they are proportional to $\frac{v_\si}{\La_\si}$, and in the limit $v_{\si}\ll \la_{\si}$ these mixing angles tend to zero. In addition, from (\ref{limmitv}) and (\ref{MWZ}), (\ref{MY}),
it follows that  $M^2_W$ is much smaller than $M^2_Y$.

\section{\label{conclus}Conclusions}
In this paper, we have modified the previous 3-3-1 model combined with discrete $S_4$ symmetry
to adapt the most recent neutrino mixing with non-zero $\theta_{13}$.
We have shown that the realistic neutrino masses and mixings can be
obtained if the two directions of the breakings $S_4\rightarrow \mathcal{K}$ and $\mathcal{K}\rightarrow Z_2$ simultaneously take
place in neutrino sector and are equivalent in size, i.e, the contributions due to $s$, $\si$ and $s'$ are comparable.
The new feature is adding a new $SU(3)_L$ anti-sextet lying in $\underline{2}$ under
 $S_4$ which can result the non-zero $\theta_{13}$ without perturbation, and consequently, the number
of Higgs multiplets required is less than
those of other models based on non-Abelian discrete symmetries and the 3-3-1 models.
The exact tribimaximal form obtained with the breaking $S_4 \rightarrow Z_3$ in charged lepton sector while
$S_4 \rightarrow \mathcal{K}$ in neutrino sector. If both the breakings
$S_4\rightarrow \mathcal{K}$ and $\mathcal{K} \rightarrow Z_2$ are taken place in neutrino sector,
the realistic neutrino spectrum is obtained without perturbation. The upper bound on neutrino
mass as well as the effective mass governing neutrinoless double beta decay at the level are presented.
 The model predicts the Dirac CP violation phase  $\delta=292.45^\circ$ in the normal spectrum (with
$\theta_{23}\neq \frac{\pi}{4}$) and $\delta=303.14^\circ$ in the inverted spectrum. We have found some regions of model parameters that can fit the experimental
data in 2014 on neutrino masses and mixing without perturbation.

\section*{Acknowledgments}
This research is funded by Vietnam National Foundation for Science
and Technology Development (NAFOSTED) under grant number 103.01-2014.51.



\appendix
\section{\label{apa}$\emph{S}_4$ group and Clebsch-Gordan coefficients}

$S_4$ is the permutation group of four objects, which is also the
symmetry group of a cube. It has 24 elements divided into 5
conjugacy classes, with \underline{1}, \underline{1}$'$,
\underline{2}, \underline{3}, and \underline{3}$'$ as its 5
irreducible representations. Any element of $S_4$ can be formed by
multiplication of the generators $S$ and $T$ obeying the relations
$S^4 = T^3 = 1,\ ST^2S = T$. Without loss of generality, we could
choose $S = (1234),\ T = (123)$ where the cycle (1234) denotes the
permutation $(1, 2, 3, 4) \rightarrow (2, 3, 4, 1)$, and (123)
means $(1, 2, 3, 4) \rightarrow (2, 3, 1, 4)$. The conjugacy
classes generated from $S$ and $T$ are\bea C_1 &:& 1 \crn C_2 &:&
(12)(34)=TS^2T^2,\ (13)(24)=S^2,\ (14)(23)=T^2S^2T\crn C_3 &:&
(123)=T,\ (132)=T^2,\ (124)=T^2S^2,\ (142)=S^2T,\crn &&
(134)=S^2TS^2,\ (143)=STS,\ (234)=S^2T^2,\ (243)=TS^2\crn C_4 &:&
(1234)=S,\ (1243)=T^2ST,\ (1324)=ST,\crn && (1342)=TS,\
(1423)=TST^2,\ (1432)=S^3\crn C_5 &:& (12)=STS^2,\ (13)=TSTS^2,\
(14)=ST^2,\crn && (23)=S^2TS,\ (24)=TST,\ (34)=T^2S\nn \eea\\
The character table of $S_4$ is given as follows

\begin{table}[ht]
\begin{center}
\begin{tabular}{@{}cccccccc@{}} \toprule
Class & $n$ & $h$ & $\chi_{\underline{1}}$ &
$\chi_{\underline{1}'}$ & $\chi_{\underline{2}}$ &
$\chi_{\underline{3}}$ & $\chi_{\underline{3}'}$ \\ \colrule
$C_1$ & 1 & 1 & 1 & 1 & 2 & 3 & 3 \\
$C_2$ & 3 & 2 & 1 & 1 & 2 & --1 & --1 \\
$C_3$ & 8 & 3 & 1 & 1 & --1 & 0 & 0 \\
$C_4$ & 6 & 4 & 1 & --1 & 0 & --1 & 1 \\
$C_5$ & 6 & 2 & 1 & --1 & 0 & 1 & --1 \\ \botrule
\end{tabular}
\end{center}
\end{table}
where $n$ is the order of class and $h$ is the order of elements
within each class. Let us note that $C_{1,2,3}$ are even
permutations, while $C_{4,5}$ are odd permutations. The two
three-dimensional representations differ only in the signs of
their $C_4$ and $C_5$ matrices. Similarly, the two one-dimensional
representations behave the same.

We will work in the basis where $\underline{3},\underline{3}'$ are
real representations whereas $\underline{2}$ is complex. One
possible choice of generators is given as follows \bea
\underline{1}&:& S=1,\hs T=1 \crn \underline{1}'
&:& S=-1,\hs T=1 \crn \underline{2} &:& S=\left(%
\begin{array}{cc}
  0 & 1 \\
  1 & 0 \\
\end{array}%
\right),\hs T=\left(%
\begin{array}{cc}
  \om & 0 \\
  0 & \om^2 \\
\end{array}%
\right)\crn \underline{3}&:& S=\left(%
\begin{array}{ccc}
  -1 & 0 & 0 \\
  0 & 0 & -1 \\
  0 & 1 & 0 \\
\end{array}%
\right),\hs T=\left(%
\begin{array}{ccc}
  0 & 0 & 1 \\
  1 & 0 & 0 \\
  0 & 1 & 0 \\
\end{array}%
\right)\crn
\underline{3}'&:& S=-\left(%
\begin{array}{ccc}
  -1 & 0 & 0 \\
  0 & 0 & -1 \\
  0 & 1 & 0 \\
\end{array}%
\right),\hs T=\left(%
\begin{array}{ccc}
  0 & 0 & 1 \\
  1 & 0 & 0 \\
  0 & 1 & 0 \\
\end{array}%
\right)\nn\eea where $\om=e^{2\pi i/3}=-1/2+i\sqrt{3}/2$ is the
cube root of unity. Using them we calculate the Clebsch-Gordan
coefficients for all the tensor products as given below.

First, let us put $\underline{3}(1,2,3)$ which means some
$\underline{3}$ multiplets such as $x=(x_1,x_2,x_3)\sim
\underline{3}$ or $y=(y_1,y_2,y_3)\sim \underline{3}$ or so on,
and similarly for the other representations. Moreover, the
numbered multiplets such as $(...,ij,...)$ mean $(...,x_i
y_j,...)$ where $x_i$ and $y_j$ are the multiplet components of
different representations $x$ and $y$, respectively. In the
following the components of representations in l.h.s will be
omitted and should be understood, but they always exist in order
in the components of decompositions in r.h.s: \bea
\underline{1}\otimes\underline{1}&=&\underline{1}(11),\hs
\underline{1}'\otimes \underline{1}'=\underline{1}(11),\hs
\underline{1}\otimes\underline{1}'=\underline{1}'(11),\crn
\underline{1}\otimes \underline{2}&=&\underline{2}(11,12),\hs
\underline{1}'\otimes \underline{2}=\underline{2}(11,-12),\crn
\underline{1}\otimes \underline{3}&=&\underline{3}(11,12,13),\hs
\underline{1}'\otimes \underline{3}=\underline{3}'(11,12,13),\crn
\underline{1}\otimes \underline{3}'&=&\underline{3}'(11,12,13),\hs
\underline{1}'\otimes \underline{3}'=\underline{3}(11,12,13),\crn
\underline{2} \otimes \underline{2} &=& \underline{1}(12+21)
\oplus \underline{1}'(12-21) \oplus \underline{2}(22,11),\crn
\underline{2}\otimes \underline{3}&=&
\underline{3}\left((1+2)1,\om (1+\om 2)2,\om^2 (1+\om^2 2)
3\right)\crn && \oplus \underline{3}'\left((1-2)1,\om (1-\om
2)2,\om^2 (1-\om^2 2) 3\right) \crn \underline{2}\otimes
\underline{3}'&=& \underline{3}'\left((1+2)1,\om (1+\om 2)2,\om^2
(1+\om^2 2) 3\right)\crn &&\oplus \underline{3}\left((1-2)1,\om
(1-\om 2)2,\om^2 (1-\om^2 2) 3\right),\crn \underline{3} \otimes
\underline{3} &=& \underline{1}(11+22+33) \oplus
\underline{2}(11+\om^2 22+ \om 33,11+\om 22+ \om^2 33) \crn
&&\oplus \underline{3}_s (23+32,31+13,12+21)\oplus
\underline{3}'_a(23-32,31-13,12-21),\crn
 \underline{3}' \otimes
\underline{3}' &=&\underline{1}(11+22+33) \oplus
\underline{2}(11+\om^2 22+ \om 33,11+\om 22+ \om^2 33) \crn
&&\oplus \underline{3}_s (23+32,31+13,12+21)\oplus
\underline{3}'_a(23-32,31-13,12-21),\crn \underline{3} \otimes
\underline{3}' &=&\underline{1}'(11+22+33) \oplus
\underline{2}(11+\om^2 22+ \om 33,-11-\om 22-\om^2 33) \crn
&&\oplus \underline{3}'_s (23+32,31+13,12+21)\oplus
\underline{3}_a(23-32,31-13,12-21),\nn\eea where the subscripts
$s$ and $a$ respectively refer to their symmetric and
antisymmetric product combinations as explicitly pointed out. We
also notice that many group multiplication rules above have
similar forms as those of $S_3$ and $A_4$ groups.

In the text we usually use the following notations, for example,
$(xy')_{\underline{3}}=
[xy']_{\underline{3}}\equiv(x_2y'_3-x_3y'_2,x_3y'_1-x_1y'_3,x_1y'_2-x_2y'_1)$
which is the Clebsch-Gordan coefficients of $\underline{3}_a$ in
the decomposition of $\underline{3}\otimes \underline{3}'$, where
as mentioned $x=(x_1,x_2,x_3)\sim \underline{3}$ and
$y'=(y'_1,y'_2,y'_3)\sim \underline{3}'$.

The rules to conjugate the representations \underline{1},
\underline{1}$'$, \underline{2}, \underline{3}, and
\underline{3}$'$ are given by \bea
\underline{2}^*(1^*,2^*)&=&\underline{2}(2^*,1^*),\hs
\underline{1}^*(1^*)=\underline{1}(1^*),\hs
\underline{1}'^*(1^*)=\underline{1}'(1^*),\crn
\underline{3}^*(1^*,2^*,3^*)&=&\underline{3}(1^*,2^*,3^*),\hs
\underline{3}'^*(1^*,2^*,3^*)=\underline{3}'(1^*,2^*,3^*),\nn\eea
where, for example, $\underline{2}^*(1^*,2^*)$ denotes some
$\underline{2}^*$ multiplet of the form $(x^*_1,x^*_2)\sim
\underline{2}^*$.

\section{\label{apt}The numbers}
In the following we will explicitly point out the lepton number
($L$) and lepton parity ($P_l$) of the model particles (notice
that the family indices are suppressed):

\begin{table}[ht]
\begin{center}
\begin{tabular}{@{}cccc@{}} \toprule
Particles & $L$ & $P_l$  \\ \colrule
$N_R$, $u$, $d$,  $\phi^+_1$,$\phi'^+_1$, $\phi^0_2$,$\phi'^0_2$,
 $\eta^0_1$,$\eta'^0_1$, $\eta^-_2$,$\eta'^-_2$
  $\chi^0_3$, $\sigma^0_{33}$, $s^0_{33}$ & 0 & 1 \\
 $\nu_L$, $l$, $U$, $D^*$, $\phi^+_3$,$\phi'^+_3$, $\eta^0_3$,$\eta'^0_3$, $\chi^{0*}_1$, $\chi^+_2$,
   $\sigma^0_{13}$,
   $\sigma^+_{23}$, $s^0_{13}$, $s^+_{23}$ & $-1$ & $-1$ \\
 $\sigma^{0}_{11}$, $\sigma^{+}_{12}$, $\sigma^{++}_{22}$,
   $s^{0}_{11}$, $s^{+}_{12}$, $s^{++}_{22}$ & $-2$ & 1\\ \botrule
\end{tabular} \label{ta2}
\end{center}
\end{table}

\section{\label{S4breaking} The breakings of $S_4$ by triplets $\underline{3}$ and $\underline{3}'$}
 
For triplets $\underline{3}$ we have the followings alignments:
\begin{itemize}
\item[(1)] The first alignment: $\langle \phi_1\rangle\neq\langle \phi_2\rangle\neq\langle \phi_3\rangle$ then $S_{4}$ is  broken into $\{1\}\equiv\{\mathrm{identity}\}$, i.e. $S_{4}$ is completely broken.
\item[(2)] The second alignment: $0\neq\langle \phi_1\rangle\neq\langle \phi_2\rangle=\langle\phi_3\rangle\neq0$ or $0\neq\langle \phi_1\rangle=\langle \phi_3\rangle\neq\langle \phi_2\rangle\neq0$ or $0\neq\langle \phi_1\rangle=\langle \phi_2\rangle\neq\langle \phi_3\rangle\neq0$ then $S_{4}$ is  broken into $Z_2$ which consisting of the elements \{$1, TSTS^2$\} or \{$1, TSS^2$\} or \{$1, S^2TS$\}, respectively.
\item[(3)] The third alignment: $\langle \phi_1\rangle=\langle
\phi_2\rangle =\langle \phi_3\rangle \neq 0$ then $S_{4}$ is  broken into $S_3$ which consisting of the elements \{$1, T, T^2, TSTS^2, STS^2, S^2TS$\}.
\item[(4)] The fourth alignment: $0=\langle \phi_2\rangle\neq\langle \phi_1\rangle=\langle \phi_3\rangle \neq 0$ or $0=\langle \phi_1\rangle\neq\langle \phi_2\rangle=\langle \phi_3\rangle \neq 0$ or $0=\langle \phi_3\rangle\neq\langle \phi_1\rangle=\langle \phi_2\rangle \neq 0$ then $S_{4}$ is  broken into $Z_2$ which consisting of the elements \{$1, TSTS^2$\} or \{$1, TSS^2$\} or \{$1, S^2TS$\}, respectively.
\item[(5)] The fifth alignment: $0=\langle \phi_2\rangle\neq\langle \phi_1\rangle\neq\langle \phi_3\rangle \neq 0$ or $0=\langle \phi_1\rangle\neq\langle \phi_2\rangle\neq\langle \phi_3\rangle \neq 0$ or $0\neq \langle \phi_1\rangle\neq\langle \phi_2\rangle\neq\langle \phi_3\rangle=0$ then $S_{4}$ is completely broken.
\item[(6)] The sixth alignment: $0\neq \langle \phi_1\rangle\neq \langle\phi_2\rangle=\langle \phi_3\rangle=0$ or $0\neq \langle \phi_2\rangle\neq \langle\phi_3\rangle=\langle \phi_1\rangle=0$
or $0\neq \langle \phi_3\rangle\neq \langle\phi_1\rangle=\langle \phi_1\rangle=0$
 then $S_{4}$ is  broken into  Klein four group $\mathcal{K}$ which consisting of the elements \{$1, S^2,TSTS^2,TST$\} or \{$1, TS^2T^2, STS^2, T^2S$\} or \{$1, T^2S^2T, ST^2, S^2TS$\}, respectively.
\end{itemize}

For triplets $\underline{3}'$ we have the followings alignments:
\begin{itemize}
\item[(1)] The first alignment: $\langle \phi'_1\rangle\neq\langle \phi'_2\rangle\neq\langle \phi'_3\rangle$ then $S_{4}$ is  broken into $\{1\}\equiv\{\mathrm{identity}\}$, i.e. $S_{4}$ is completely broken.
\item[(2)] The second alignment: $0\neq\langle \phi'_1\rangle\neq\langle \phi'_2\rangle=\langle\phi'_3\rangle\neq0$ or $0\neq\langle \phi'_1\rangle=\langle \phi'_3\rangle\neq\langle \phi'_2\rangle\neq0$ or $0\neq\langle \phi'_1\rangle=\langle \phi'_2\rangle\neq\langle \phi'_3\rangle\neq0$ then $S_{4}$ is  broken into $\{1\}\equiv\{\mathrm{identity}\}$, i.e. $S_{4}$ is completely broken.
\item[(3)] The third alignment: $\langle \phi'_1\rangle=\langle
\phi'_2\rangle =\langle \phi'_3\rangle \neq 0$ then $S_{4}$ is broken into $Z_3$ that consists of the elements \{$1, T, T^2$\}.
\item[(4)] The fourth alignment: $0=\langle \phi'_2\rangle\neq\langle \phi'_1\rangle=\langle \phi'_3\rangle \neq 0$ or $0=\langle \phi'_1\rangle\neq\langle \phi'_2\rangle=\langle \phi'_3\rangle \neq 0$ or $0=\langle \phi'_3\rangle\neq\langle \phi'_1\rangle=\langle \phi'_2\rangle \neq 0$ then $S_{4}$ is  broken into $Z_2$ which consisting of the elements \{$1, T^2S$\} or \{$1, TST$\} or \{$1, ST^2$\}, respectively.
\item[(5)] The fifth alignment: $0=\langle \phi'_2\rangle\neq\langle \phi'_1\rangle\neq\langle \phi'_3\rangle \neq 0$ or $0=\langle \phi'_1\rangle\neq\langle \phi'_2\rangle\neq\langle \phi'_3\rangle \neq 0$ or $0\neq \langle \phi'_1\rangle\neq\langle \phi'_2\rangle\neq\langle \phi'_3\rangle=0$ then $S_{4}$ is completely broken.
\item[(6)] The sixth alignment: $0\neq \langle \phi'_1\rangle\neq \langle\phi'_2\rangle=\langle \phi'_3\rangle=0$ or $0\neq \langle \phi'_2\rangle\neq \langle\phi'_3\rangle=\langle \phi'_1\rangle=0$
or $0\neq \langle \phi'_3\rangle\neq \langle\phi'_1\rangle=\langle \phi'_1\rangle=0$
 then $S_{4}$ is  broken into a four-element subgroup generated by a four-cycle, which consisting of the elements \{$1, S, S^2,S^3$\} or \{$1, TST^2, ST, TS^2T^2$\} or \{$1, TS, T^2ST, T^2S^2T$\}, respectively.
\end{itemize}

\end{document}